\documentclass[12pt,titlepage]{article}
\usepackage{graphicx}
\usepackage{endfloat}
\usepackage{amsfonts,amsmath,amssymb}
\usepackage{mychicago}
\usepackage{subfigure}
\usepackage{genetics_manu_style}
\renewcommand{\eqref}[1]{Equation~\ref{#1}}
\newcommand{\mni}{V}

\title{Rate of adaptation in sexuals and asexuals: \\ A solvable model of the Fisher-Muller effect}

\author{Su-Chan Park\thanks{Department of Physics, The Catholic University of Korea, Bucheon 420-743, Republic of Korea} and Joachim Krug\thanks{Institut  
f\"ur Theoretische Physik, Universit\"at zu K\"oln, 50937 K\"oln, Germany}}

\renewcommand{\CorrespondingAddress}{
    Department of Physics \\
    The Catholic University of Korea\\
    43 Jibong-ro, Wonmi-gu,\\ 
    Bucheon 420-743, Republic of Korea\\ 
    Tel : +82-2-2164-4524\\
    e-mail : spark0@catholic.ac.kr
    \vfill}
\renewcommand{\RunningHead}{Rate of adaptation in sexuals and asexuals}
\renewcommand{\CorrespondingAuthor}{Su-Chan Park}
\renewcommand{\KeyWords}{advantage of recombination, clonal
  interference, speed of adaptation}
\begin{document}

\maketitle 
 
\begin{abstract}
The adaptation of large asexual populations is hampered by the
competition between independently arising beneficial mutations in different 
individuals, which is known
as clonal interference. In classic work, Fisher and Muller proposed
that recombination provides an evolutionary advantage in large
populations by alleviating this competition. Based on recent progress
in quantifying the speed of adaptation in asexual populations
undergoing clonal interference, we present a detailed analysis of the
Fisher-Muller mechanism for a model genome consisting of two loci with
an infinite number of beneficial alleles each and multiplicative
(non-epistatic) fitness effects. We solve the deterministic, infinite
population dynamics exactly and show that, for a particular, natural
mutation scheme, the speed of adaptation in sexuals is twice as large
as in asexuals. This result is argued to hold for any nonzero value
of the rate of recombination. Guided by the infinite population result
and by previous work on asexual adaptation, we postulate an expression
for the speed of adaptation in finite sexual populations that agrees
with numerical simulations over a wide range of population sizes and
recombination rates. The ratio of the sexual to asexual adaptation
speed is a function of population size that increases in the clonal
interference regime and approaches 2 for extremely large populations.
The simulations also show that the imbalance between the numbers of 
accumulated mutations at the two loci is strongly suppressed even by a small amount 
of recombination.
The generalization of the model to an arbitrary number $L$ of loci is
briefly discussed. 
If each offspring samples the alleles at each locus from the gene pool of the whole population
rather than from two parents,
the ratio of the sexual to asexual adaptation speed is approximately equal 
to $L$ in large populations. A possible realization of this scenario is the reassortment
of genetic material in RNA viruses with $L$ genomic segments. 
\end{abstract}
The evolutionary advantage of sex remains one of the most intriguing
puzzles in evolutionary biology~\cite{K1993,dVE2007,O2009}.
Many hypotheses have been suggested explaining why sexual reproduction
is widespread in nature despite apparent disadvantages such as the two-fold 
cost of sex~\cite{M1978}. Well-known examples are the deterministic mutation 
hypothesis~\cite{K1988}, the Fisher-Muller
mechanism~\cite{F1930,M1932,CK1965} and  
Muller's ratchet~\cite{M1964,F1974}, to name only a few. 
These three hypotheses are applicable when the
fitness landscape in question has certain specific features. 
Specifically, the deterministic mutation hypothesis requires 
deleterious mutations to be synergistically epistatic, while 
the Fisher-Muller (FM) mechanism 
as well as Muller's ratchet can explain the advantage of 
sex if epistasis is negligible.

Theoretical analyses of the effect of epistasis on the speed of Muller's ratchet 
have concluded that it practically stops operating when epistasis is synergistic ~\cite{CMC1993,K1994,J2008}.
Furthermore, recent experimental analyses of empirical fitness 
landscapes seem to indicate that a particularly strong form of
epistasis termed sign epistasis~\cite{WWC2005} is quite common~\cite{WDDH2006,dVPK2009,FKdVK2011,Szendro2013}. 
Sign epistasis generally implies that the fitness 
landscape is rugged. On a rugged fitness landscape sex can be
detrimental, even without taking into account the two-fold cost of sex,  
in that sexual populations, unlike the corresponding asexual populations, 
cannot escape from local fitness peaks~\cite{CK1965,Eshel1970,dVPK2009,PK2011}. 

Although research on empirical fitness landscapes has been growing
substantially in recent years, 
it is still practically infeasible to reliably determine genotypic
fitness on a genome-wide scale [but see \citeN{Kouyos2012}]. 
Because of the small sizes of most empirical fitness landscapes that
have so far been constructed experimentally, the implications of sign epistasis for long term
evolution remain unclear.  
At the same time experimental evidence
in favor of the FM mechanism has also accumulated~\cite{C2002,C2007}. 
For these reasons further quantitative analysis of the advantage of sex 
in the \textit{absence} of epistasis remains a worthwhile endeavor, and
we will pursue this approach in the present contribution. 

The essence of the FM mechanism is the competition between independently arising
beneficial mutations, termed clonal interference, which slows down the adaptation of large asexual
populations~\cite{GL1998,MGME1999,W2004,KO2005,PK2007,FND2008,Sniegowski2010,Schiffels2011}. The concept of clonal interference has played
an important role in interpreting the behavior observed in laboratory
selection experiments~\cite{LRST1991,LT1994,Barrick2009}, and has  
also been invoked in explaining the population-size dependence of
evolutionary predictability in rugged fitness landscapes~\cite{JPK2011,Szendro2013b}. 
Although in its original formulation clonal 
interference theory neglects the occurrence of secondary beneficial 
mutations within a growing clone \cite{GL1998,G2001}, in general the
coexistence of multiple beneficial mutations cannot be neglected in
large populations~\cite{PK2007}. In the following we will therefore use the term clonal
interference in a wider sense than originally conceived, in that two clones with different
numbers of beneficial mutations can compete with each other for fixation.

Much recent theoretical work has focused on obtaining accurate
quantitative estimates of the speed of adaptation in the presence of
clonal interference for the simple situation of an unlimited supply of
beneficial mutations that act independently on fitness, without
epistatic interactions [see~\citeN{PSK2010} for review].   
It turns out that the population dynamics in this regime is well described
by a traveling wave moving at constant speed along a one-dimensional
fitness space. The traveling wave picture was first established for
the case when the beneficial selection coefficient is the same for all 
mutations~\cite{TLK1996,RWC2003,DF2007,BRW2008,RBW2008} and recently extended to the
more realistic case of selection coefficients drawn from a continuous
effect size distribution~\cite{GRBHD2012,Fisher2013}; see also~\citeN{B1999} and
references therein for a traveling wave picture of adaptation in changing environments.

In natural populations it is unlikely that the traveling
wave picture persists forever. Apart from the assumed absence of
epistatic interactions, there are two main features
that lead to a breakdown of this picture in long-term evolution. 
First, a fluctuating environment generally makes the fitness landscape change
with time. In a time-dependent situation it is problematic
to compare absolute fitnesses of two individuals living on different landscapes
and, accordingly, adaptation is measured through the relative fitness increase or its time-integrated
form termed fitness flux~\cite{Mustonen2010}. 
Second, even if the fitness landscape remains constant for a very 
long time, the indefinite supply of beneficial mutations appearing at constant
rate cannot be a good approximation in the real world.
For example, in long-term evolution experiments the speed
of adaptation usually slows down~\cite{LT1994,Barrick2009}, which is attributed to 
the decreasing supply of beneficial mutations. In this context, the
house-of-cards model, in which fitness values are assigned randomly to
genotypes, could provide a more realistic
description~\cite{K1978,PK2008}. In the framework of this model one
cannot however explain the advantage of sex, because the fitness of a recombinant
genotype is uncorrelated with the parental fitnesses and therefore 
beneficial mutations cannot accumulate through recombination. 

Although the non-epistatic model with an infinite supply of beneficial
mutations is of limited validity, it can provide a reasonable
approximation when a population undergoes 
a severe environmental change, as is often the case at the
beginning of an evolution experiment. At the same time this setting 
is conceptually simple and allows for detailed (if approximate) mathematical
analysis. In the present paper, we therefore build upon the recent
line of work on asexual populations undergoing clonal interference and
add to it a minimal yet realistic recombination scheme. 
Specifically, we consider a sexual population model with 
two genetic loci, each of which can acquire infinitely many beneficial mutations. 
For simplicity we 
assume that epistasis is absent both between and within loci. 
Upon reproduction, the offspring receives one locus from each parent with
probability $r$ and both loci from a single parent with probability $1-r$. 
A possible biological realization of this kind of facultatively sexual reproduction
is the assortment of genetic material in RNA viruses with two genomic segments,
where the parameter $r$ reflects the probability of co-infection and is governed
by the multiplicity of infection, the ratio of viruses to the number of infected cells 
\cite{SLH2011}. In this context it is natural to consider the 
generalization of the model to $L$ loci, which will be described in DISCUSSION. 

We first analyze the infinite population dynamics of the two-locus model, obtaining exact
expressions for the speed of adaptation in the limiting cases of 
zero and maximal recombination rates (asexuals vs. obligate sexuals). When the selection
coefficient of beneficial mutations is the same at both loci and at
most one mutation may occur per generation and individual, the speed
of adaptation for obligate sexuals is twice that of asexuals, a result
that we argue holds for any positive recombination rate. Based on this
observation we conjecture that for finite populations the speed of adaptation in sexuals 
is approximately equal to the sum of the speeds of the two loci, each of which receives
half of the supply of beneficial mutations. Denoting the speed of adaptation by $v_s$ for 
sexuals and by $v_a$ for asexuals, and the genome-wide beneficial mutation rate by $U$, 
the conjectured relation reads
\begin{equation}
\label{Eq:twice}
v_s(U) \approx 2 v_a(U/2).
\end{equation}
This relation has two important implications. First,
provided the asexual speed of adaptation increases more slowly than linear with the mutation rate $U$,
as is clearly the case in the presence of clonal interference, sexuals are at an advantage in the sense
that $v_s(U) > v_a(U)$. In fact, since the asexual speed becomes almost independent of the mutation supply 
rate for very large populations~\cite{PSK2010}, there is a two-fold advantage of sex in this regime. 
Second, the precise theoretical estimates for the speed of adaptation in asexuals that have been developed
in recent work translate through Equation \ref{Eq:twice} into explicit expressions for the sexual speed of adaptation in our model. 
In RESULTS we present a detailed comparison of Equation \ref{Eq:twice} to finite population 
simulations, finding good agreement already for small recombination rates.  
In DISCUSSION we address the consequences of relaxing some of the assumptions of our model,
describing in particular a possible extension of the model to more than two loci, 
and place our work into
the context of related studies. 

\section{Models} 
We consider a sexual or asexual population of haploid individuals 
in discrete generations. The population size is denoted by $N$ and assumed to 
be constant.  As a reproduction scheme we employ the Wright-Fisher 
model~\cite{F1930,W1931}, the prototypical model of discrete, non-overlapping 
generations. Since our main concern is how recombination
affects the speed of adaptation, we 
assume that all mutations are beneficial. This naturally leads us to 
study evolution in the framework of the infinite-sites
model~\cite{K1969}; 
otherwise back mutations of beneficial mutations, which are deleterious by 
definition, should appear with nonzero probability. Furthermore, we assume no epistasis among mutations, which will be reflected by the multiplicative fitness assignment. 
As a minimal model with the above properties, we study 
an evolving population with only two loci under selection. Each locus is
assumed to have infinitely many sites. We assume an initially 
homogenous population and the fitness of the initial genome, or 
wild-type, is set to unity.

In line with the assumption of multiplicative fitness effects, the
fitness of an individual that has $n_i$ mutations at 
locus $i$ compared to the wild-type is $\exp(n_1 s_1 + n_2 s_2)$.
Without loss of generality we take $s_2\ge s_1$.
Note that two genotypes with the same number of mutations at each locus are not 
necessarily the same though both have the same fitness 
$\exp(s_1 n_1 + s_2 n_2)$. 
Since we are only interested in how fast mean fitness increases and not
in the genealogy, all genotypes with the same number of mutations at each locus will be 
treated as if they were the same.

The population evolves in the following way.
Let $f_t(n_1,n_2)$ denote the frequency of all genotypes with $n_1$ mutations at the first locus
and $n_2$ mutations at the second locus at generation $t$.
At $t=0$, the population is homogeneous with $f_0(0,0) =1$.
By selection, the frequency at generation $t+1$ on average will change to be
\begin{equation}
f_{t}^s(n_1,n_2) = \frac{e^{s_1 n_1 + s_2 n_2}}{\bar w_t} f_t(n_1,n_2),
\label{Eq:S}
\end{equation}
where 
\begin{equation}
\bar w_t \equiv \sum_{n_1,n_2} e^{s_1 n_1 + s_2 n_2} f_t(n_1,n_2)
\end{equation}
is the average fitness of the population at generation $t$. 

Mutation can also change the frequency of genotypes. 
The probability that an offspring is hit by $m_1$ mutations at the first locus and 
$m_2$ mutations at the second locus will be denoted by $g_0(m_1,m_2)$. Here we implicitly assume that the mutation 
probability is not affected by the genetic background.
To be concrete, $g_0(0,0)$ is the probability that neither locus is mutated, $g_0(1,0)$ is the probability that a 
mutation occurs at the first locus, but not at the second, and so on. 
In most of our analysis, we will assume that $g_0(0,0) + g_0(1,0) + g_0(0,1) = 1$, 
which reflects that only single-site mutations can occur.
The frequency change due to both selection and mutation is
\begin{eqnarray}
\nonumber
f_{t}^\mu(n_1,n_2) &=& \sum_{m_1,m_2\ge 0} g_0(n_1-m_1,n_2-m_2)f_t^s(m_1,m_2)\\
&=& \sum_{m_1,m_2\ge 0}g_0(n_1-m_1,n_2-m_2) \frac{e^{s_1m_1+s_2m_2}}{\bar w_t} f_t(m_1,m_2),
\label{Eq:SM}
\end{eqnarray}
where $g_0(x,y)$ with at least one negative argument should be
understood to be 0. We further assume that mutation 
does not have any preference for a certain locus, that is,
$g_0(m_1,m_2) = g_0(m_2,m_1)$ for any pair of $m_1$ and $m_2$. 

After selection and mutation, two randomly chosen parents mate and beget
an offspring. Let 
$R(n_1,n_2 | k_1,k_2;l_1,l_2)$ denote the probability that
the resulting progeny of two individuals with respective genotypes
$(k_1,k_2)$ and $(l_1,l_2)$ has the genotype $(n_1,n_2)$. 
To be specific, we set $(0\le r \le 1)$
\begin{equation}
R(n_1,n_2|k_1,k_2;l_1,l_2) =
\begin{cases} 
(1-r)/2 &\text{if }n_1 = k_1, n_2=k_2,\\
(1-r)/2 &\text{if }n_1 = l_1, n_2=l_2,\\
r/2 &\text{if }n_1 = k_1, n_2 = l_2,\\
r/2 &\text{if }n_1 = l_1, n_2 = k_2,\\
0&\text{otherwise,}
\end{cases}
\label{Eq:WFdyn}
\end{equation}
which means that with probability $1-r$ the two loci
 of the offspring in question are
inherited solely from a single parent which is selected with probability 1/2 and with probability $r$ the offspring
inherits one locus from one parent and the other from the other parent.
When $r=0$, an offspring inherits all genotypes from a single parent, 
so we will call the case with $r=0$ asexuals.
On the other hand, when $r=1$, an offspring inherits alleles from both parents, 
so we will call the case with $r=1$  obligate sexuals.
In this sense, the case with $0<r<1$ can be regarded as facultatively sexual populations.


Since the probability that the randomly chosen parents have genotypes
$(k_1,k_2)$ and $(l_1,l_2)$ is $f_{t}^\mu(k_1,k_2)f_{t}^\mu(l_1,l_2)$,
the mean frequency after selection, mutation, and recombination is
\begin{eqnarray}
\nonumber
f_{t}^r(n_1,n_2) 
&=& \sum_{k_1,k_2,l_1,l_2} R(n_1,n_2|k_1,k_2;l_1,l_2) f_{t}^\mu(k_1,k_2)f_{t}^\mu(l_1,l_2)\\
&=& (1-r) f_t^\mu(n_1,n_2) + r f_t^{(1)}(n_1) f_t^{(2)}(n_2),
\label{Eq:SMR}
\end{eqnarray}
where
\begin{eqnarray}
f_t^{(1)}(n_1) = \sum_{n_2} f_t^{\mu}(n_1,n_2),\quad
f_t^{(2)}(n_2) = \sum_{n_1} f_t^{\mu}(n_1,n_2),
\end{eqnarray}
are marginal frequency distributions of genotypes after the selection and mutation
steps with $n_1$ mutations at locus 1 and $n_2$ mutations at locus 2, 
respectively. 

Finally, the actual population distribution
at generation $t+1$ is determined by multinomial sampling using
$f_t^r(n_1,n_2)$ in \eqref{Eq:SMR} with the restriction that
the population size is $N$. For simulations, we employ the algorithm explained
by \citeN{PK2007} [see also \citeN{PSK2010} for simulations of
extremely large populations]. 

The speed of adaptation, or shortly speed, is defined as the rate of
increase of the log mean fitness,
\begin{equation}
v \equiv \lim_{t\rightarrow \infty} \frac{\langle \ln \bar w_t \rangle }{t},
\end{equation}
where $\langle \ldots \rangle$ denotes an average over independent 
realizations of evolution with the same parameters.
In the following, we mainly focus on the dependence of speed on 
parameters such as the population size, the mutation probability per generation, the selection 
coefficient of a single mutation, and the recombination probability. 

\section{Results}
\subsection{Infinite populations:}
Although the infinite population limit cannot be reached in  
real biological populations for the model we are studying~\cite{PSK2010},
it does provide some insight into the adaptation dynamics of finite populations.
Furthermore, the deterministic nature of the infinite population dynamics
renders an analytic approach feasible. We therefore begin our
discussion with the evolutionary dynamics of infinite populations.
Detailed derivations and generalizations of the results presented here 
can be found in APPENDIX A.

As shown in APPENDIX A, the advantage of sex in infinite populations
depends on the exact form of mutation probability distribution $g_0$. 
However, as will be demonstrated later in DISCUSSION, the form of $g_0$ does not
affect the speed of populations with biologically relevant size as long
as the mutation probability is small. In the following 
we employ the simple mutation scheme 
\begin{equation}
g_0(0,0) = 1 - U,\quad g_0(1,0) = g_0(0,1) = \frac{U}{2},
\label{Eq:mut}
\end{equation}
which does not allow for multiple-site mutations. 
In this case, the speed for asexuals, $v_a$ and for obligate sexuals, $v_s$, 
are found to be (see APPENDIX A)
\begin{equation}
v_a = \max(s_1, s_2) = s_2,\quad v_s = s_1 + s_2.
\label{Eq:INF_speed}
\end{equation}
This result can be understood as follows:
for obligate sexuals ($r=1$), the two loci are unlinked and, thus, each locus 
evolves independently with mutation probability $U/2$. Since, regardless of the
actual value of $U\neq 0$, the contributions
from each locus are $s_1$ and $s_2$, respectively,  
the total speed $v_s$ is the sum of these two. For asexuals, 
clonal interference 
prohibits accumulation of the weaker beneficial effect $s_1$, so the
speed is determined solely by the larger beneficial effect $s_2$.
\eqref{Eq:INF_speed} is also valid when $s_1 = s_2$. In this case,
$v_s$ is twice as large as $v_a$, that is, a two-fold advantage of sex, 
which is the maximum effect of sex in the two-locus model.
When we study the adaptation dynamics of finite populations, we will
set $s_1 = s_2 = s$ to maximize the advantage of sex.

Although we only found the speed exactly for the cases $r=0$ and $r=1$,
we now argue that the asymptotic speed does not depend on $r$ 
provided $r > 0$ for any mutation scheme. Let $\ell_1(t)$ ($\ell_2(t)$) denote the maximum number of 
mutations at locus 1 (locus 2) accumulated up to generation $t$:
\begin{equation}
\ell_1(t) \equiv \text{max}\left \{ n \left | \sum_m f_t(n,m)\right . \neq 0\right \},\;
\ell_2(t) \equiv \text{max}\left \{ m \left | \sum_n f_t(n,m)\right . \neq 0\right \}.
\label{Eq:def_lead}
\end{equation}
This definition can be used for finite populations as well, and is
closely related to the \textit{lead} of the fitness distribution
considered in the traveling wave approach to asexual adaptation \cite{DF2007,PSK2010,Fisher2013}.  
Within our general mutation scheme with homogeneous initial conditions,
$\ell_i(t) = M t$ for infinite populations,
where $M$ is the largest possible number of sites
that can be mutated at one locus in a single mutation event.
Hence the frequency $f_t(Mt,Mt)$ of genotypes with $M t$ mutations at each locus at generation $t$ is nonzero due to recombination,
though it can be extremely small. 

Now assume that the speed $v_s(r,N=\infty)$ for $0<r<1$
is strictly smaller than $M (s_1+s_2)$. Then, with time $t$,
the ratio of the detectable largest fitness to the mean fitness
increases as $\exp(( M (s_1+s_2) -v_s) t)$. 
Thus, at some $t$, the relative fitness of the genotypes with $M t$ mutations at
each locus to the mean fitness becomes extremely large, 
which eventually results in
an abrupt increase of frequency of these genotypes in one generation.
Accordingly, $\bar w_t$ becomes of the order of $\exp( M( s_1 + s_2) t)$,
and in the long run the speed becomes $M(s_1 + s_2) $ for any $r>0$. 

\begin{figure}
\includegraphics[width=\textwidth]{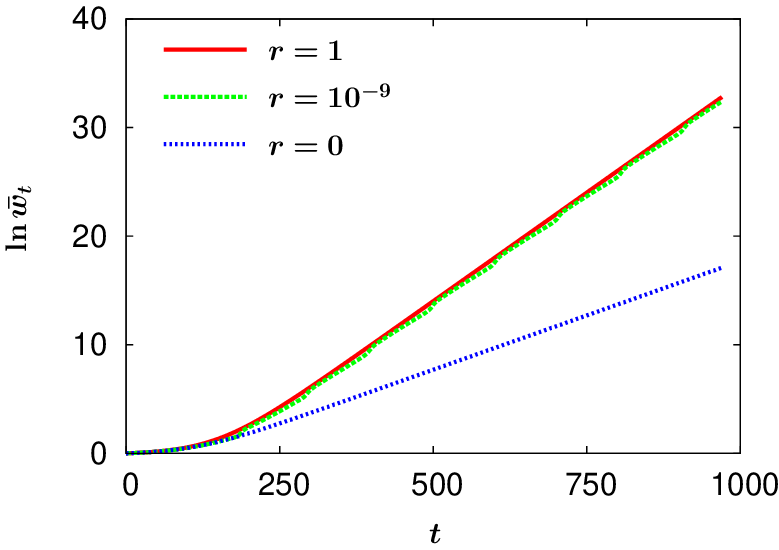}
\caption{\label{Fig:rs} Log-mean fitness $\ln \bar w_t$ of the infinite
  population model
as a function of time for
$r=0$, $r=10^{-9}$, and $r=1$ (from bottom to top) with $U=0.1$ and $s=0.02$.
As argued in the text, the speed does not depend on $r$ once $r$ is nonzero.
}
\end{figure}

In the above discussion, we argued that the speed does not depend on $r$ once $r$ is nonzero. On the other hand, if $r$ is very small, the whole population 
behaves almost like an asexual population for quite some time. Hence, the 
abrupt jump of fitness mentioned above should be observable. To see this phenomenon, 
we studied the deterministic evolution numerically, using
the mutation scheme of \eqref{Eq:mut} with $U=0.1$ and $s_1=s_2=0.02$. 
In Fig. \ref{Fig:rs}, we show how the mean fitness behaves with time for
$r=0$, $r=10^{-9}$, and $r=1$. Even for the minute recombination rate
of $r=10^{-9}$, the mean fitness closely follows the $r=1$ curve,
however with some oscillations. To elucidate the origin of this
behavior we need to consider how the frequency distribution changes with time. 

\begin{figure}
\includegraphics[width=\textwidth]{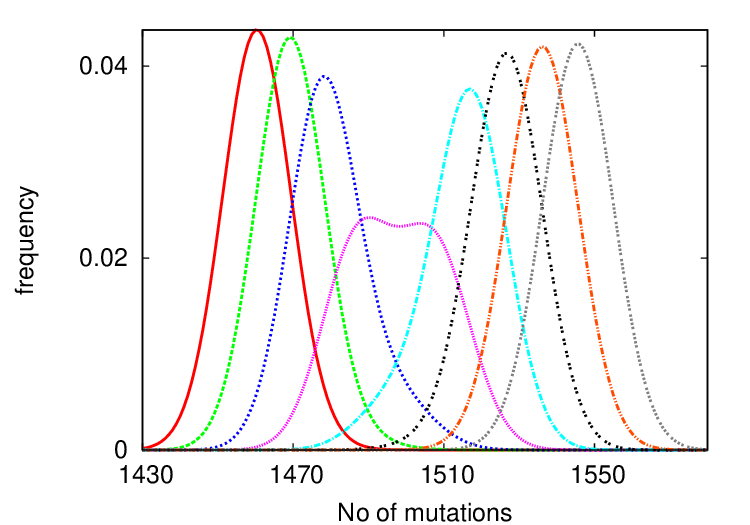}
\caption{\label{Fig:config} Frequency distribution of the
  total number of
mutations for the infinite population model at generations $895, 900, 905, \ldots, 
930$ (left to right) with
parameters $r=10^{-9}$, $U=0.1$, and $s=0.02$.}
\end{figure}

In the asexual case ($r=0$) the frequency distribution over the number of mutations
is well described by a Gaussian~\cite{PSK2010}.
Furthermore, the frequency distribution of the obligately sexual population with $r=1$ 
should also be well described by a Gaussian, because the generating function
is just the product of two generating functions of asexual evolution
(see \eqref{Eq:part}).
However, for $0<r\ll 1$, the Gaussian may not be a good approximation.
In Fig.~\ref{Fig:config} we depict the time evolution of the frequency distribution for 
$r=10^{-9}$. Clearly the frequency distribution
cannot be approximated by a Gaussian traveling wave. Moreover, the shape of 
the distribution changes with time, which implies that there is 
no time-independent steady state. Rather, the distribution behaves like a `breathing traveling wave' in that the 
behavior seen in Fig.~\ref{Fig:config} repeats periodically.
In Supporting Information, one can find an animation showing the breathing traveling wave.
The time when two peaks become comparable in Fig.~\ref{Fig:config} 
corresponds to the abrupt jump of mean fitness alluded to
above. Further mathematical analysis of this phenomenon seems interesting, but we will
not pursue it here because it is hardly observable in real, finite populations. 

\subsection{Finite populations:}
We mentioned before that many analytic approaches
have been developed to find an expression for the speed of adaptation
in large asexual populations~\cite{RWC2003,DF2007,BRW2008,RBW2008}. \citeN{PSK2010} summarized these developments and 
compared simulation results with the proposed analytic expressions. 
The approximation of \citeN{RBW2008} turned out to be quite accurate 
in a wide range of parameters. 
The only disadvantage of this approach is that the speed is obtained as an implicit function of $N$ (see below). 
In this section, 
we will find a mathematical formula for the speed of adaptation in sexual populations,
using both the suggested formula for asexuals and 
the results for the infinite population dynamics in the previous section. 

For an infinite population, as shown in APPENDIX A,
the precise form of the mutational probability distribution $g_0(k_1,k_2)$ 
affects the speed.
However, for plausible values of the
mutation rate and the selection coefficient such infinite population
effects become observable only for unrealistically large populations \cite{PSK2010}, see DISCUSSION for a detailed argument. 
In the following we therefore use \eqref{Eq:mut} and set
$s_1 = s_2 = s$ for the reasons mentioned previously. This implies that 
at most one mutation can occur per individual in each generation, and all mutations
have the same selective effect $s$.

\begin{figure}
\includegraphics[width=\textwidth]{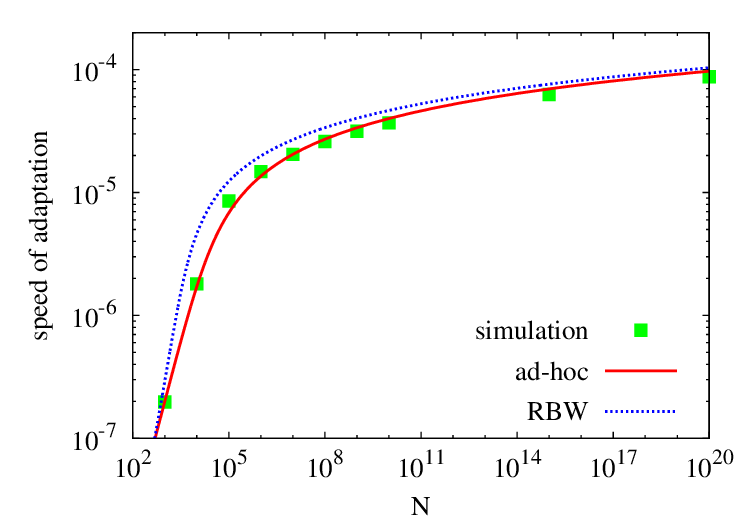}
\caption{\label{Fig:adhoc} Speed of adaptation of finite asexual populations
as a function of $N$ on a double logarithmic scale for $U=10^{-6}$ and
$s=0.01$. 
The numerical solutions of \eqref{Eq:asex_speed}  and
\eqref{Eq:asex_speed_adhoc} are drawn for comparison with the
simulation data. As anticipated, the {\it ad-hoc} modification
(\eqref{Eq:asex_speed_adhoc}) provides a more accurate estimate.}
\end{figure}

We begin with a discussion of the speed for asexual populations.
As was illustrated by \citeN{PSK2010}, the speed for the asexual version of our 
model ($r=0$) is well approximated by the implicit equation
\begin{equation}
\ln N \approx \frac{v_a^\mathrm{RBW}}{2 s^2} \left ( \ln^2 \frac{v_a^\mathrm{RBW}}{e U s} + 1 
\right ) - \ln \sqrt{\frac{s^3 U}{v_a^\mathrm{RBW} \ln(v_a^\mathrm{RBW}/(Us))}},
\label{Eq:asex_speed}
\end{equation}
where the subscript $a$ in $v_a$ refers to the asexual population, 
the superscript RBW refers to the authors of \citeN{RBW2008}, and 
$e \approx 2.718182$ is the base of the natural logarithm.
Since the approximation of the fitness distribution by a continuous traveling wave was used to derive \eqref{Eq:asex_speed}, 
it should not be surprising that the discrepancy between theory and 
simulation becomes relatively large when the size of population is small enough
to realize the strong-selection weak-mutation (SSWM) regime, where the
population is mostly monomorphic.
Based on this observation, there is room for improvement of the approximation in an {\it ad-hoc} way
as follows: First we note that the first term 
in \eqref{Eq:asex_speed} is dominant when the speed is high 
and the second term is dominant when the speed is low. 
Thus, when the population size is small, we can neglect the first term. 
In the SSWM regime, two consecutive fixations of beneficial mutations can be considered independent,
so the speed can be estimated as the mean number of fixed mutations per generation times the selection coefficient of the fixed mutation.
Since the fixation probability of a beneficial mutation with selection coefficient $s$ is approximately $2 s$
and all beneficial mutations have the same effect in our model,
the speed in the SSWM regime is $v_a = N U \times 2 s \times s = 2 N U s^2$.
Using the speed in the SSWM regime, we modify \eqref{Eq:asex_speed} as
\begin{equation}
\ln N \approx \frac{v_a}{2 s^2} \left ( \ln^2 \frac{v_a}{e U s} + 1 
\right ) + \ln \frac{v_a}{2s^2 U}
\label{Eq:asex_speed_adhoc}
\end{equation}
which keeps the large speed behavior unchanged and enforces the
SSWM result for small speeds.
In Fig.~\ref{Fig:adhoc} we show that \eqref{Eq:asex_speed_adhoc}
provides a more accurate approximation to the speed obtained from simulations with
$U=10^{-6}$ and $s=0.01$ than \eqref{Eq:asex_speed}.

\begin{figure}
\includegraphics[width=\textwidth]{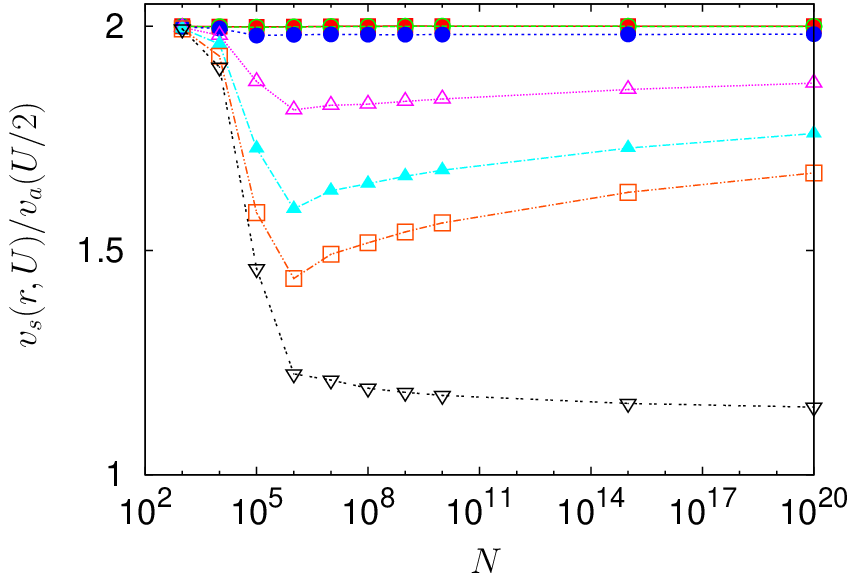}
\caption{\label{Fig:tworel} Ratio of the sexual adaptation speed,
  $v_s(r,U)$, to the asexual speed at half mutation rate, $v_a(U/2)$,
  as a function of population size $N$.
Recombination rates are $r=0$ (empty reverse triangle), $10^{-5}$ (empty square), $10^{-4}$ (filled
triangle), $10^{-3}$ (empty triangle),$ 10^{-2}$ (filled circle),$ 10^{-1}$ (empty circle), and 1 (filled square)
from bottom to top, and $U=10^{-6}$ and
$s=0.01$ are used throughout. 
The scaling relation in \eqref{Eq:ansatz} predicts that $v_s(r,U)/v_a(U/2) =
2$. Note that two datasets for $r=0.1$ (empty circle) and $r=1$ (filled square) are indiscernible.
}
\end{figure}

Now we move on to the speed of sexual populations. At first, let us start
from the case of $r=1$ whose infinite population limit
allows for an exact solution. As we show in APPENDIX A, 
the evolutionary dynamics of an infinite population with $r=1$ can be
viewed as the independent evolution of each locus with the marginal mutation
probability $\tilde g_0(k)$. That is, we can divide the evolutionary dynamics
into two independent asexual populations with reduced
mutation probability and the speed of the sexual
population 
is obtained by simply adding the speeds of these two virtual asexual
populations. Within the mutation scheme given by \eqref{Eq:mut} with 
selection coefficients $s_1 =s_2 =s$ this implies that 
\begin{equation}
v_s\left ( r=1,U  \right ) = 2 v_a\left (U/2\right )
\label{Eq:ansatz}
\end{equation}
for sufficiently large populations.
Interestingly, \eqref{Eq:ansatz} is trivially valid  
in the SSWM regime where the speed is linear in $U$ and 
$v(r,U) \approx 2 N U s^2$ irrespective of $r$.
Since \eqref{Eq:ansatz} accurately estimates the speed for
very small and very large populations, it is likely that
\eqref{Eq:ansatz} is a good approximation for any population size.
Indeed, as we show in Fig.~\ref{Fig:tworel}, $v_s(r=1,U)$ is 
well approximated by twice $v_a(U/2)$ for any population size.
The parameters we have used in these simulations are $U=10^{-6}$ and $s=0.01$.
With the help of \eqref{Eq:asex_speed_adhoc}, we may thus approximate the speed
$v_s$ of the obligately sexual population as
\begin{equation}
\ln N \approx \frac{v_s}{4 s^2} \left ( \ln^2 \frac{v_s}{e U s} + 1 
\right ) + \ln \frac{v_s}{2s^2 U}.
\label{Eq:sex_speed}
\end{equation}

It is clear that for $0<r\ll 1$ there should be a regime
where \eqref{Eq:ansatz} cannot approximate the speed accurately.
To see this deviation, we simulated populations with various $r$ (Fig.~\ref{Fig:tworel}).
It turns out that \eqref{Eq:ansatz} is still a good approximation
for $r \ge 10^{-2}$. In particular, the speed
for $r=0.1$ is hardly discernible from that for $r=1$ for all population sizes.
The deviation starts to be significant for $r = 10^{-3}$. 
For comparison, we also plot $v_s(r=0,U)/v_a(U/2)$ or equivalently
$v_a(U)/v_a(U/2)$ in Fig.~\ref{Fig:tworel}, which should approach to 1
in the infinite population limit \cite{PSK2010}. 
For $N\ge 10^6$, where $NU\ln(Ns)$ becomes larger than 1, 
$v_s(r,U)/v_a(U/2)$ starts to increase though very slowly and 
$v_s(r,U)$ becomes significantly larger than $v_s(r=0,U) = v_a(U)$.
Note that for asexual populations clonal interference sets in around
$N U \ln (Ns) \sim 1$ \cite{W2004,PSK2010}. That is, as soon as clonal interference becomes
relevant, even a small amount of recombination 
leads to a significant speedup of adaptation, in agreement with the FM mechanism. 

\begin{figure}
\includegraphics[width=\textwidth]{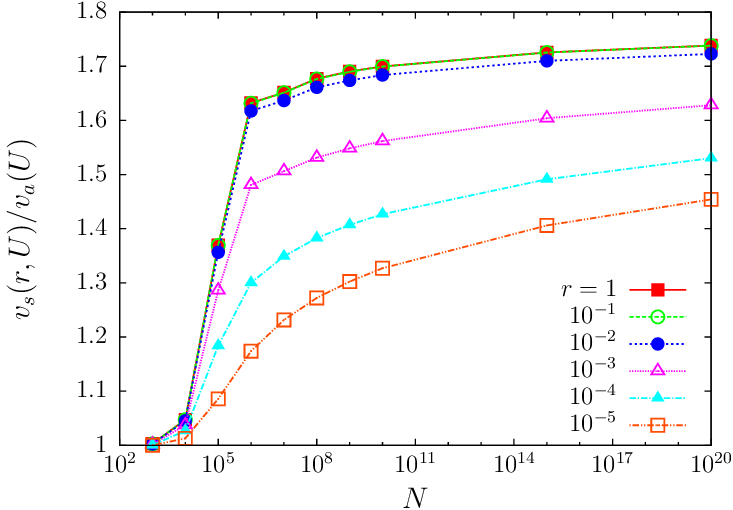}
\caption{\label{Fig:vsvaU} Ratio of sexual to asexual speed of adaptation, $v_s(r,U)/v_a(U)$,
  as a function of population size $N$ on a semi-logarithmic scale. 
Recombination rates are $r=10^{-5}$, $10^{-4}$, $10^{-3}$, $10^{-2}$, $10^{-1}$, and 1
from bottom to top, and $U=10^{-6}$ and $s=0.01$ are used as before. 
As in Fig.~\ref{Fig:tworel}, the two datasets for $r=1$ and $r=0.1$ are hardly discernible.
}
\end{figure}

To display the FM effect more clearly, we depict $v_s(r,U)/v_a(U)$ vs $N$ in Fig.~\ref{Fig:vsvaU}.
The fact that the ratio $v_s(r,U)/v_a(U)$ continues to rise
monotonically with $N$ for all cases with $r > 0$ in Figs.~\ref{Fig:tworel} and \ref{Fig:vsvaU} is consistent with the 
two-fold advantage predicted by the infinite population analysis. 



\begin{figure}
\includegraphics[width=\textwidth]{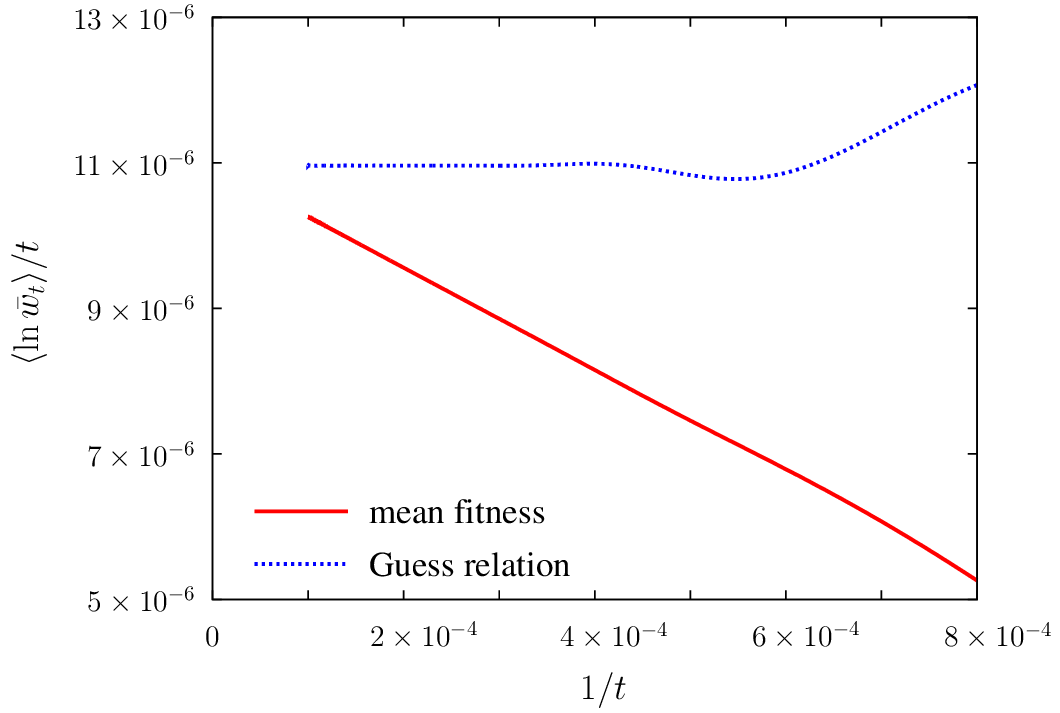}
\caption{\label{Fig:Guess} Numerical verification of the Guess
  relation. The figure compares the logarithmic
  mean fitness divided by generation time, $\langle \ln \bar w_t \rangle /t$,
to the right hand side of \eqref{Eq:Guess}. 
The data are obtained from simulations with $r=10^{-3}$, $U=10^{-6}$ and
$s=0.01$. For $t\rightarrow \infty$, both curves intersect the ordinate
at the same point, which equals the asymptotic speed of adaptation. 
}
\end{figure}

When measuring the speed of adaptation in our simulations, a useful
consistency check was provided by the Guess relation
\begin{equation}
v = U s + \left \langle \frac{1}{N} 
\sum_{i=1}^N (\chi_i - 1) \ln \chi_i \right \rangle_\text{stat},
\label{Eq:Guess}
\end{equation}
where $\chi_i$ is the relative fitness of $i$-th individual in the infinite
time limit,
\begin{equation}
\chi_i = \lim_{t\rightarrow \infty} \frac{w_i(t)}{\bar w(t)},
\end{equation} 
and $\langle \cdot \rangle_\text{stat}$ signifies an average over
the stationary measure of $\chi_i$. 
\eqref{Eq:Guess} was originally established for asexual populations
undergoing discrete generation (WF) dynamics~\cite{G1974T,G1974A}. 
In APPENDIX B, we prove that the relation holds for sexuals as well,
and in Fig.~\ref{Fig:Guess}, we numerically confirm its validity.
The two terms on the right hand side of \eqref{Eq:Guess} represent the
increase in population fitness due to mutation and selection,
respectively. Recombination affects the speed of adaptation only
indirectly through its effect on the relative fitnesses $\chi_i$. 
Note that the Guess relation should hold
even if one uses discrete-time, overlapping generation models
such as the Moran model.

Finally, we analyze the difference in the number of beneficial mutations
acquired by the two loci. We quantify this difference as
\begin{equation}
\mni \equiv  \lim_{t\rightarrow \infty} 
\frac{\left \langle\left ( \ell_1(t) - \ell_2(t) \right )^2\right \rangle}{t},
\label{Eq:skewness}
\end{equation}
where $\ell_1$ and $\ell_2$ are defined in \eqref{Eq:def_lead}.
We will refer to $\mni$ as the mutation number imbalance (MNI).
To discern the MNI of asexuals from that of sexuals,
we will add subscripts $a$ and $s$, for
asexual and sexual populations, respectively.
In infinite populations each locus accumulates the same number of mutations, hence this study is meaningful only for finite populations. 

\begin{figure}
\includegraphics[width=\textwidth]{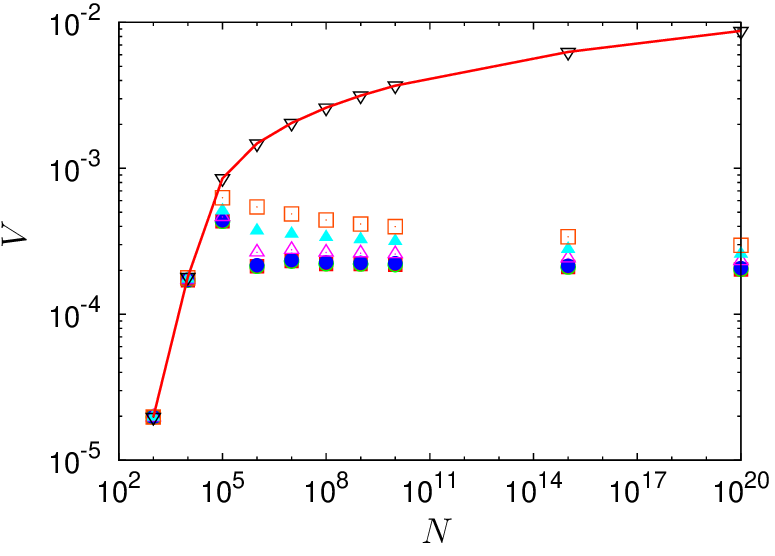}
\caption{\label{Fig:skewa} The mutation number imbalance (MNI)
  $\mni$ vs. population size $N$ for 
$r=0$ (empty reverse triangle), $10^{-5}$ (empty square), $10^{-4}$ (filled
triangle), $10^{-3}$ (empty triangle),$ 10^{-2}$ (filled circle),$ 10^{-1}$ (empty circle), and 1 (filled square)
from top to bottom. Other parameter values are $U=10^{-6}$ and
$s=0.01$.
The symbols for $r \ge 10^{-2}$ essentially overlap.
For comparison, the analytic prediction $\mni_a = v_a/s$ for asexuals
is drawn as a line.
}
\end{figure}

For asexual populations, an approximation for $\mni$ can be obtained
by comparing the \textit{origination processes} at the two loci, which count
the mutations that are present in some individuals of the population at time $t$  
and that are destined to eventually go to fixation \cite{G1993,Gillespie_Book,PK2007}. 
Denoting the number of such mutations at locus $i$ by $k_i(t)$, we
assume that 1) the difference between $k_i(t)$ and the
lead $\ell_i(t) \geq k_i(t)$ remains bounded in the long time limit, 
2) the total number of mutations $M_F = k_1 + k_2$ in the origination process increases at the same rate
as the mean number of mutations, $M_F(t) \approx v_a t/s$ for large $t$, and 3) each new mutation appearing
in the origination process chooses one of two loci with equal probability. Assumptions 1) and 2) reflect the
existence of a steady state and have been verified in simulations
\cite{PK2007}, and assumption 3) is a consequence of the symmetry between 
the two loci. By assumption 3),
the probability that there are $m_1$ mutations at locus 1 and $m_2 =
M_F(t) -m_1$ mutations at locus 2 is given by
\begin{equation}
P(m_1,t) \approx \binom{M_F(t)}{m_1} \left ( \frac{1}{2} \right )^{M_F(t)}.
\end{equation}
Since the mean of $m_1$ is $\langle m_1 \rangle = M_F/2$ and its
variance is $\langle ( m_1 - \langle m_1 \rangle )^2 \rangle = M_F/4$,
we can calculate $\mni_a$, invoking the assumption 1), as
\begin{equation}
\mni_a \approx \frac{\left \langle \left (M_F(t) - 2 m_1 \right )^2 \right \rangle}{t}
= \frac{4}{t} \left \langle \left (m_1 - \langle m_1 \rangle \right )^2 \right \rangle
\rightarrow \frac{v_a }{s}.
\end{equation}
In Figure \ref{Fig:skewa}, we compare $\mni_a$ to $v_a/s$ for $U=10^{-6}$
and $s=0.01$, which shows an excellent agreement.

Recombination changes the behavior of the MNI 
substantially. As can be seen in Figure~\ref{Fig:skewa}, once the population size is in the regime of 
clonal interference the MNI decreases abruptly, then 
remains almost constant for a wide range of population sizes.
Even a small amount of recombination efficiently equalizes any major
fitness difference between the two loci by creating competitively
superior recombinants in which both loci have high fitness. 


\section{DISCUSSION} 

The Fisher-Muller mechanism for the evolutionary advantage of sex is
based on the slowing down of asexual adaptation due to clonal
interference, which is alleviated by the recombination of high fitness 
genotypes. While much recent theoretical work has been devoted to
quantifying the speed of adaptation in asexuals, the speedup that can
be achieved through recombination has been explicitly addressed only
in a few studies (see below). In the present article we
take a step in this direction by providing a detailed analysis of a
simple, yet biologically meaningful model in which recombination occurs
between two loci, each of which can harbor an unlimited number of linked
beneficial mutations. Our analysis shows that the advantage of sex becomes significant
in the parameter regime where clonal interference plays an
important role in asexual populations. In our two-locus model, 
the adaptation speed of sexual populations is about twice as large
as that of the corresponding asexual populations for a 
wide range of recombination rate. 
In the remainder of this section we discuss the robustness of our results to relaxing
some of the assumption in our model, in particular the neglect of multiple and recurrent 
mutations. We then describe a possible extension of the model to $L $ loci and discuss
its relevance to the adaptation of RNA viruses with multiple genetic segments. Finally, 
we briefly compare our findings to related previous work.

\subsection{Multiple-site mutations:}
In most of the analysis and simulations presented above we have 
assumed that only single-site mutations can occur in an individual each generation. 
Since mutations are replication errors that may occur at multiple sites in an independent fashion, 
a more realistic assumption would be that the probability for $n$ mutations to arise in 
one individual is of order $U^n$, where $U$ is the probability of a single-site mutation. 
In the following we argue that allowing for multiple-site
mutations does not significantly affect our results for 
the speed of adaptation for any biologically plausible population size, provided 
$U$ is small. 

In the SSWM regime, $N U \ll 1$,
multiple-site mutations obviously cannot contribute to the adaptation dynamics and 
\eqref{Eq:twice} remains valid. On the other hand, in the infinite 
population limit the speed of adaptation 
strongly depends on the form of the mutation probability $g_0(m_1,m_2)$
(see \eqref{Eq:speed_inf_a} and \eqref{Eq:speed_inf_s}). To be concrete, we adopt the
mutation scheme of \eqref{Eq:mut2} which allows for mutations at up to two sites, with two-site mutations
occurring with probability $U^2$.  The infinite population analysis then predicts
that $v_s = v_a$, hence \eqref{Eq:twice} must break down beyond some
characteristic population size $N_c$. 

\begin{figure}
\includegraphics[width=\textwidth]{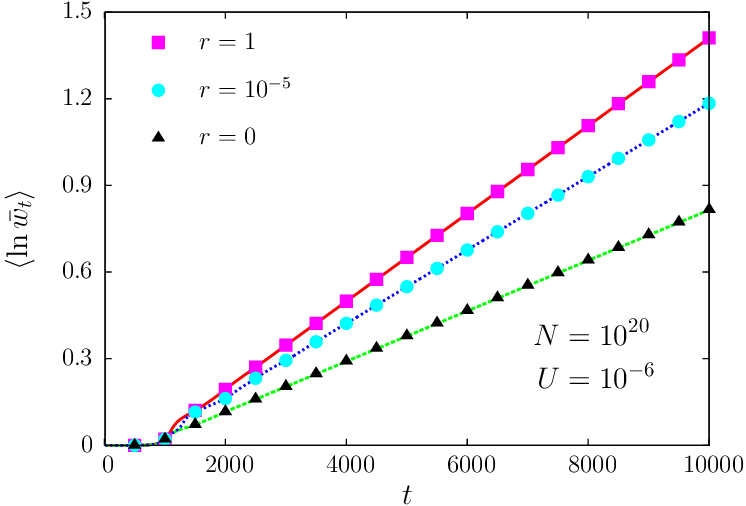}
\caption{\label{Fig:multimut} Mean logarithmic fitness $\langle \ln \bar w_t \rangle$ vs time $t$
in the presence (symbols) and absence (lines) of two-site mutations. The mutation schemes employed
in the two cases are given in \eqref{Eq:mut2} and \eqref{Eq:mut}, respectively. 
The population size is $N=10^{20}$ and the probability for a single mutation is $U=10^{-6}$. 
The two data sets are indistinguishable, which implies that 
multiple-site mutations do not play any role.}
\end{figure}

A first guess about $N_c$ invokes the criterion for the onset of clonal interference.
Since clonal interference among single-site mutations becomes important when 
$N U \ln N \ge 1$ \cite{GL1998,W2004,PSK2010}, clonal interference among clones with two-site
mutations would become important when $N U^2 \ln N \ge 1$. Thus, for $U=10^{-6}$ as 
assumed in our simulations, the effect of multiple-site mutations should be observable for $N \gg 10^{10}$.
To check the validity of this argument, we simulated the model for $N=10^{20}$ and $U=10^{-6}$ using
the mutation scheme of \eqref{Eq:mut2} and compare the results to those presented
previously assuming that only single-site mutations are possible, see Fig.~\ref{Fig:multimut}.
Contrary to the above expectation, no detectable difference is observed. In fact, we could not
observe any significant difference even for $N = 10^{100}$, even though in that case 
about $10^{88}$ double mutants occur in every generation (results not shown).

The reason for the failure of the above criterion is that multiple-site mutations can affect
the speed of adaptation only if they occur among the offspring of the fittest individuals 
in the population. 
Within the traveling wave picture of asexual adaptation, these individuals reside 
in the so-called stochastic edge which governs the rate of advance of the entire
population~\cite{DF2007,BRW2008,RBW2008,GRBHD2012,Fisher2013}, while mutations occuring in the bulk of the
traveling wave are wasted by clonal interference. 
If the total number of offspring of the stochastic edge class per generation is much smaller than $U^{-2}$,
a mutant offspring of the edge class is most likely to have a single-site mutation and, accordingly,
single-site mutations should play a dominant role in the advance of the stochastic edge.

To find $N_c$, consider a large asexual population such that the selection coefficient of the fittest 
class, $\tilde s$, relative to the mean fitness is large and loss of the stochastic edge by genetic drift is unlikely. 
If this is not the case, the edge almost always starts from a single individual with few offspring and, in turn, 
multiple-site mutations cannot affect the speed for the reason given above. When only single-site mutations can occur, 
$\tilde s=\ln (1/U)$ for an infinite 
population and the frequency of individuals in this maximum fitness class with $t$ mutations is of order 
$\exp\left [-\ln^2 U / (2 s) \right ]$~\cite{PSK2010}. Thus for a population with size 
$N \ge \exp\left [\ln^2 U / (2 s)  \right ]$ the fittest class is occupied by at least one individual at all times 
and the traveling wave reaches the deterministic speed limit $v_a =
s$; for this finite population $\tilde s$ is also $\ln (1/U)$. The mean number of
offspring of an individual in the fittest class is of order $e^{\tilde s} = 1/U$, so that
on average one of the offspring will gain an additional
mutation, securing the advance of the wave at maximum
speed. Correspondingly, when double mutations are allowed and occur at
rate $U^2$, the number of individuals in the fittest class investigated above must be of
order $1/U$ to ensure that one double mutant can be created with high probability from this class.
We therefore conclude that 
multiple-site mutations will affect the speed of adaptation only if 
\begin{equation}
\label{Eq:Nc}
N \ge N_c = U^{-1} \exp\left [\ln^2 U / (2 s)  \right ].
\end{equation} 
Since $N_c \approx 10^{4150}$ for $U = 10^{-6}$ and $s = 0.01$, multiple site mutations
cannot change the outcome for any biologically reasonable population size. 
This implies that, in contrast to the inifinite population model, the dynamics of finite populations
are remarkably robust with regard to changes in the mutation scheme.

Although the above conclusion has been arrived at only by analyzing asexual populations, multiple-site mutations
in sexual populations cannot affect the speed for any biologically relevant population size because
the fittest class of each locus still has a small number of individuals; see also Fig.~\ref{Fig:multimut} for
numerical support. 

\subsection{Finite number of sites:} We next discuss the implications of relaxing our assumption 
that each of the two loci carries an infinite number of sites at which beneficial mutations can
occur. If the number of sites is finite, there is a nonzero probability that the same site
will be hit multiple times. Two cases must be distinguished. If a beneficial mutation that
was previously lost by genetic drift or clonal interference arises a second time, its effect
will not be different from that of a new mutation in the infinite sites model, and in that
sense such recurrent mutations are already accounted for in our analysis. On the other hand,
if a site at which a beneficial mutation has been fixed is hit again, it constitutes 
a deleterious mutation. As long as such events are rare, the deleterious mutations will quickly
be purged by natural selection. However, in the long run this leads to a depletion of the
(finite) supply of beneficial mutations and causes the rate of adaptation to slow down
in sexuals as well as asexuals, a regime that is beyond the scope of our study. 

When the number of sites is finite, the Fisher-Muller effect thus gives rise to 
a transient advantage of sex that has been studied quantitatively by \citeN{KO2005}.
They find that the speedup due to recombination
is maximal when all beneficial mutations have the same selective strength and becomes
less pronounced when different mutations have different strengths.  
Within our infinite sites model this aspect could be addressed by allowing
for a distribution of mutational effects instead of a single selection coefficient $s$.

\subsection{More than two loci:}
It is natural to surmise that the factor of two arises in our model
because we are considering two loci, and that the speed increase
should in general be proportional to the number of loci. Indeed, 
this turns out to be true if we use a `communal' recombination scheme where 
the gene of each locus is collected from the whole population rather than
from the two parents, once the genome of the offspring is 
constructed  by recombination. In a three-locus model with the `communal' 
recombination scheme, the frequency distribution
of next generation is sampled from 
\begin{equation}
f_t^r(n_1,n_2,n_3) = (1-r) f_t^\mu(n_1,n_2,n_3) + 
r f_t^{(1)}(n_1) f_t^{(2)}(n_2) f_t^{(3)} (n_3),
\end{equation}
where $f_t^{(i)}(n_i)$ is the marginal frequency distribution 
for having $n_i$ mutations at locus $i$ after the (deterministic)
selection and mutation steps (compare to MODELS).
It is a straightforward extension of the calculation in APPENDIX A
to show that the infinite population dynamics for $r=1$ is again divided
into three independent evolutions of each locus with marginal
mutation probabilities just as in the two-locus case. In general,
if we consider a model system with $L$ loci within the communal
recombination scheme mentioned above, the evolution 
is a superposition of $L$ independent evolutions at
each locus, and there is an $L$-fold advantage of sex in the infinite population.

\begin{figure}
\includegraphics[width=\textwidth]{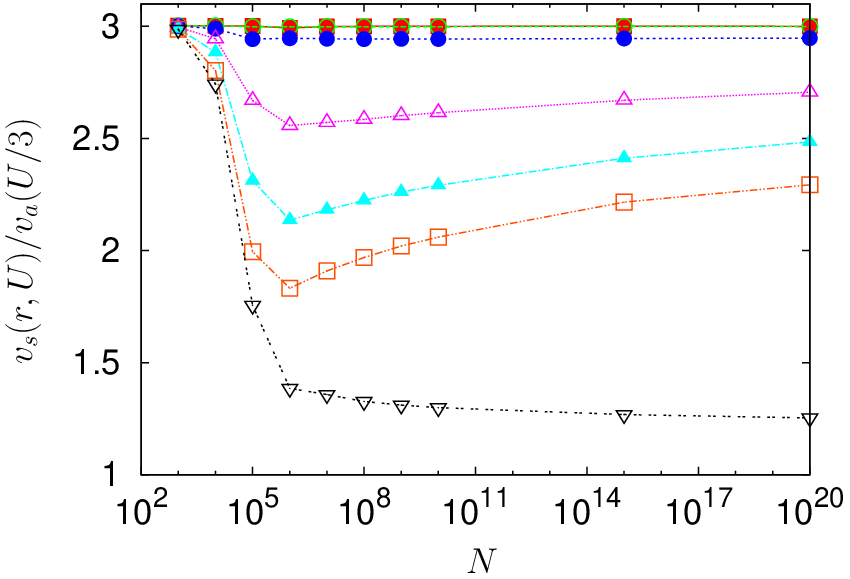}
\caption{\label{Fig:threerel}  Ratio of the sexual adaptation speed in
  the three-locus model, $v_s(r,U)$, to the asexual speed $v_a$ at mutation
  rate $U/3$ as a function of population size $N$.
Recombination rates are $r=0$ (empty reverse triangle), $10^{-5}$ (empty square), $10^{-4}$ (filled
triangle), $10^{-3}$ (empty triangle),$ 10^{-2}$ (filled circle),$ 10^{-1}$ (empty circle), and 1 (filled square)
from bottom to top, and $U=10^{-6}$ and
$s=0.01$ are used throughout. 
The scaling relation in \eqref{Eq:three_sex} predicts that
$v_s(r,U)/v_a(U/3) = 3$. Note that two datasets for $r=0.1$ (empty circle) and $r=1$ (filled square) are indiscernible.
}
\end{figure}

\begin{figure}
\includegraphics[width=\textwidth]{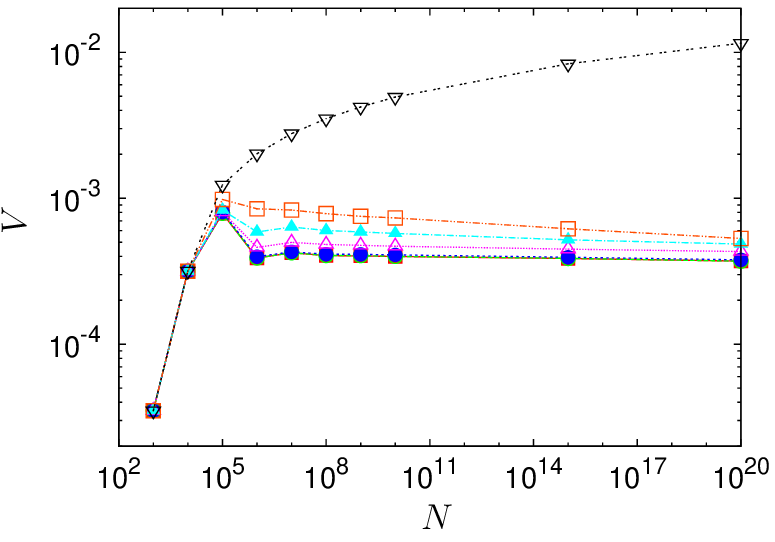}
\caption{\label{Fig:threemni} Mutation number imbalance (MNI)
for the three-locus model with recombination rates  
$r=0$ (empty reverse triangle), $10^{-5}$ (empty square), $10^{-4}$ (filled
triangle), $10^{-3}$ (empty triangle),$ 10^{-2}$ (filled circle),$ 10^{-1}$ (empty circle), and 1 (filled square)
from top to bottom. Other parameter values are $U=1.5\times 10^{-6}$ and $s=0.01$.
The symbols for $r \ge 10^{-2}$ essentially overlap.
}
\end{figure}

To see if this $L$-fold advantage persists for finite populations,
we performed simulations of the three-locus model. As in the two-locus
case, we expect that 
\begin{equation}
v_s(r,U) = 3 v_a(U/3)
\label{Eq:three_sex}
\end{equation}
for sufficiently large $r$. 
Indeed, we observe that the simulations 
are consistent with \eqref{Eq:three_sex} for a wide range of parameter
values. In Fig.~\ref{Fig:threerel}, we depict $v_s(r,U)/v_a(U/3)$ as a function
of $N$ for $U=1.5\times 10^{-6}$ and $s=0.01$ with varying $r$.
As in the  two-locus model, the advantage of sex becomes significant
when clonal interference is important in the corresponding asexual
populations.
We also studied the mutation number imbalance in the three-locus model.
We slightly modify the definition of the MNI as the difference between
the maximum and minimum numbers of accumulated mutations at all loci,
which reduces to the definition of \eqref{Eq:skewness} for the two-locus model.
In Fig.~\ref{Fig:threemni}, we depict the MNI for the three-locus
model with $U=1.5\times 10^{-7}$ and $s=0.01$ for various $r$.
Like the MNI of the two-locus model, the MNI for the asexuals
increases with $N$ while slowly decreasing for sexuals.
Again, the qualitative difference between sexuals and asexuals becomes
significant in the regime where clonal interference is important.

\subsection{Genetic reassortment in RNA viruses:} The communal recombination scheme described above
arises naturally in RNA viruses with $L$ genomic segments which are reassorted during the 
coinfection of a single cell by several viruses \cite{SLH2011}. Since the degree of reassortment
can be controlled via the multiplicity of infection, this class of systems offers the 
opportunity to test hypotheses concerning the evolutionary advantage of recombination through
the direct comparison between sexual and asexual populations \cite{Chao1990,MGME1999,Poon2004}. 
Of particular interest in the context of our work is a study by \citeN{Turner1998} which aimed
to test the Fisher-Muller mechanism by measuring the rate of fitness increase for 
the $\phi 6$ bacteriophage in the presence and absence of reassortment. Surprisingly, the asexual
populations were found to adapt faster because a possible advantage of sexuals is more than offset
by an additional cost due to intrahost competition during coinfection. If this complication 
could be avoided through an appropriate experimental design, RNA viruses would provide
a suitable framework for experimentally testing the predictions of the present paper.  

\subsection{Relation to previous studies and outlook:}
The quantitative analysis of the speed of evolution of sexual populations 
compared to that of asexual populations, when both evolve on the same non-epistatic
fitness landscape with the same beneficial mutation rate per genome,
has a long history \cite{CK1965,MS1968,CK1969,MS1971,F1974,MS1976,M1978,KO2005}. When the number
of accessible beneficial mutations is finite, the relevant quantity is
the time for all beneficial mutations to be fixed. In this context,
\citeN{MS1971} analyzed the fixation time for sexual and asexual
populations evolving on a fitness landscape with $L$ loci under selection.
Each locus has two alleles, one of which confers a beneficial fitness
effect in a non-epistatic fashion.  For sexual populations, the linkage 
among loci was assumed weak. Using a rough approximation,
\citeN{MS1971} argued that the time for completing evolutionary changes in asexual populations
is $L$ times longer than that in sexual populations for 
sufficiently large population size, which implies an $L$-fold 
advantage of sex similar to what we found in our study (see also \citeN{MS1976}). 
While the analysis by \citeN{MS1971} is fairly crude and (as conceded by the
author) actually not consistent with the simulation results presented in
the same paper, the conclusion that the advantage of sex becomes stronger with 
an increasing number of loci under selection is in qualitative
agreement with our results, as well as with the related work of
\citeN{KO2005}. 

Recent studies of the speed of sexual populations in the context of the FM 
mechanism have mostly focused on the case with an infinite supply of beneficial 
mutations~\cite{NSF2010,RC2010,WB2012}, exploiting the mathematical
progress in treating the spreading of beneficial mutations as a 
Gaussian traveling wave~\cite{TLK1996,RWC2003,DF2007,RBW2008,PSK2010,GRBHD2012,Fisher2013}.
\citeN{RC2010} studied how recombination speeds up adaptation when
there is standing variation of beneficial mutations. 
\citeN{NSF2010} studied the speed of adaptation of large facultatively
sexual populations, starting from a monomorphic state.
Similar to our results, \citeN{NSF2010} found a regime of intermediate
recombination rates where the speed increases logarithmically with 
population size, however with a prefactor that varies quadratically
with $r$. Although \citeN{RC2010} and 
\citeN{NSF2010} investigated the adaptation dynamics of 
sexual populations with an (effectively) infinite supply of beneficial
mutations, their results cannot be directly compared to ours. 
This is because the model genomes of \citeN{RC2010} and
\citeN{NSF2010} assume no or weak linkage between beneficial mutations,  
whereas in our model mutations in the same locus are tightly linked. 
Stated differently, unlike our model which
allows for an infinite number of possible beneficial alleles per locus, each 
locus in the models cited above has only two possible alleles. 
A related study with an explicit genetic map was recently presented by 
\citeN{WB2012}. In future work, it may be of interest to consider 
models in which the number of linked sites per locus, the number of loci and 
the rate and mode of recombination can all be varied independently, and the 
different limiting cases considered in these earlier studies and in the present
work can be explored in a unified setting.     

The two-locus genome considered in this paper can be viewed as a
simple example of a modular genomic architecture, where recombination
occurs between modules but not within a module. \citeN{WWW2011} have
pointed out that such a modular structure induces a strong
benefit for sexual reproduction when there is sign epistasis within
the modules and different modules contribute independently to
fitness. Another promising avenue for future research would therefore be 
to extend our approach to include a tunable degree of epistasic interactions within the loci. 
Following \citeN{WWW2011}, such interactions should affect not only
the speed of adaptation but also the set of genotypes that can be
reached at all by the population.


\section{Acknowledgments}
Support by Deutsche Forschungsgemeinschaft within SFB 680 
{\it Molecular Basis of Evolutionary Innovations} is gratefully acknowledged. 
In addition, S.-C.P. acknowledges the support by the Basic Science Research 
Program through the National Research Foundation of Korea (NRF) 
funded by the Ministry of Education, Science and Technology 
(Grant No. 2011-0014680) and the Catholic University of Korea, Research
Fund, 2012. We thank Pleuni Pennings and an anonymous reviewer for helpful comments on 
an earlier version of the manuscript. 
\bibliographystyle{mychicago} 
\bibliography{ParkKrug}

\renewcommand{\theequation}{A\arabic{equation}}
\setcounter{equation}{0}
\section{APPENDIX A: Infinite population dynamics for asexuals
  \lowercase{($r=0$)} and obligate sexuals \lowercase{($r=1$)}}
When the population size is infinite, 
the frequency of genotypes with $n_i$ mutations at locus $i$ 
at generation $t+1$, $f_{t+1}(n_1,n_2)$, is equal to $f_t^r(n_1,n_2)$
as given in \eqref{Eq:SMR} due to the law of large numbers. 
For the deterministic dynamics, the method of (moment) generating functions 
has been successfully applied to models with non-epistatic fitness 
landscapes~\cite{J1999,MBF2003,PK2007}, and we employ this method in this APPENDIX.

Let $F_t(z_1,z_2)$ denote the generating function for the frequency distribution
at generation $t$, which is defined as
\begin{equation}
F_t(z_1,z_2) \equiv \sum_{n_1,n_2} z_1^{n_1} z_2^{n_2} f_t(n_1,n_2).
\end{equation}
Since the fitness landscape is multiplicative, the mean fitness at generation 
$t$ can be found from $F_t$ through
\begin{equation}
\bar w_t = F_t(e^{s_1},e^{s_2}).
\end{equation}
Likewise, we introduce the generating function for $f_t^s$ in \eqref{Eq:S},
which is obtained from $F_t$ according to
\begin{equation}
F_t^s(z_1,z_2) \equiv \sum_{n_1,n_2} z_1^{n_1} z_2^{n_2} f_t^s(n_1,n_2)
= \frac{F_t(e^{s_1} z_1, e^{s_2} z_2)}{F_t(e^{s_1},e^{s_2})}.
\end{equation}
Since $f_t^\mu$ in \eqref{Eq:SM} is the convolution of $g_0$ and $f_t^s$, the 
generating function for $f_t^\mu$ is the product of $F_t^s$ and $G_0(z_1,z_2)$,
where $G_0$ is the generating function for mutation probability $g_0$ defined
as
\begin{equation}
G_0(z_1,z_2) \equiv \sum_{k_1,k_2} z_1^{k_1} z_2^{k_2} g_0(k_1,k_2).
\label{Eq:GF_g0}
\end{equation}
Using that $f_{t+1}$ is the same as $f_t^r$ for infinite populations,
we obtain an iterative evolution equation for $F_t$ that reads
\begin{eqnarray}
\label{Eq:GF}
F_{t+1}(z_1,z_2) &=& (1-r) F_t^{\mu}(z_1,z_2) + r F_t^{\mu}(1,z_2)F_t^{\mu}(z_1,1)\\
&=&  (1-r) G_0(z_1,z_2) \frac{F_t(z_1 e^{s_1},z_2 e^{s_2})}{F_t( e^{s_1}, e^{s_2})}
+  r  \widetilde G_0(z_1)\frac{F_t(z_1 e^{s_1}, e^{s_2})}{F_t( e^{s_1}, e^{s_2})}
\widetilde G_0(z_2)\frac{F_t( e^{s_1},z_2 e^{s_2})}{F_t( e^{s_1}, e^{s_2})} ,\nonumber
\end{eqnarray}
where $F_t^\mu(z_1,z_2)= G_0(z_1,z_2) F_t^s(z_1,z_2) $ is the generating function of $f_t^\mu$ and 
\begin{equation}
\widetilde G_0(z)\equiv G_0(z,1) = \sum_{k_1,k_2} z^{k_1} g_0(k_1,k_2)= 
G_0(1,z) =\sum_{k_1,k_2} z^{k_2} g_0(k_1,k_2) 
\label{Eq:Gen_margin}
\end{equation}
can be regarded as the generating function for the marginal mutation
probability 
\begin{equation}
\tilde g_0(k) = \sum_{m} g_0(k,m)= \sum_{m} g_0(m,k).
\end{equation}
Note that we are using the symmetry $g_0(k_1,k_2) = g_0(k_2,k_1)$ introduced
earlier, but the generalization to asymmetric $g_0$ is straightforward

For $r=0$, \eqref{Eq:GF} can be solved by iterating 
backwards until $t=0$, that is~\cite{PK2007}
\begin{eqnarray}
F_{t}(z_1,z_2) &=& G_0(z_1,z_2) \frac{F_{t-1}(z_1 e^{s_1},z_2 e^{s_2})}{F_{t-1}( e^{s_1}, e^{s_2})}
= G_0(z_1,z_2) \frac{G_0(z_1 e^{s_1},z_2 e^{s_2}) }{G_0(e^{s_1},e^{s_2})} \frac{F_{t-2}(z_1 e^{2s_1},z_2 e^{2s_2})}{F_{t-2}( e^{2s_1}, e^{2s_2})}\nonumber\\
&=& \frac{F_0(z_1e^{ts_1},z_2 e^{ts_2})}{F_0(e^{ts_1},e^{ts_2})} \prod_{\tau=0}^{t-1} \frac{ G_0(z_1e^{s_1\tau},z_2e^{s_2\tau})}{ G_0(e^{s_1 \tau},e^{s_2\tau})}
=\prod_{\tau=0}^{t-1} \frac{ G_0(z_1e^{s_1\tau},z_2e^{s_2\tau})}{ G_0(e^{s_1 \tau},e^{s_2\tau})}
\end{eqnarray}
where we have used $F_0(z_1,z_2) = 1$ for the homogeneous initial condition.
Thus the mean fitness at generation $t$ is
\begin{equation}
\bar w_t = G_0(e^{s_1t},e^{s_2t} ).
\label{Eq:AI}
\end{equation}

For $r=1$, \eqref{Eq:GF} suggests that each locus evolves independently and, 
in turn, that the generating function is the product of two functions such as
\begin{equation}
F_{t}(z_1,z_2) = \widetilde F^1_t(z_1) \widetilde F^2_t(z_2),
\label{Eq:part}
\end{equation}
which can be considered the absence of linkage between two locus, or
linkage equilibrium.
With the above ansatz, we can find an evolution equation for
$\widetilde F^i_t(z)$ ($i=1$ or 2) from \eqref{Eq:GF},
\begin{equation}
\widetilde F^i_{t+1}(z) = \widetilde G_0(z) \frac{\widetilde F^i_t(ze^{s_i})}{\widetilde F^i_t(e^{s_i})},
\label{Eq:evol_asex}
\end{equation}
which is exactly the evolution equation for an asexual population with
marginal mutation probability $\tilde g_0$.
Hence the solution of \eqref{Eq:evol_asex} is
\begin{equation}
\widetilde F^i_t(z) = \prod_{\tau=0}^{t-1} \frac{\widetilde G_0(ze^{s_i\tau})}{\widetilde G_0(e^{s_i \tau})},
\label{Eq:LEsol}
\end{equation}
where we have again used the homogeneous initial condition $\widetilde F^i_0(z) =1$.
One can easily check that \eqref{Eq:part} with
$\widetilde F^i_t(z)$ in \eqref{Eq:LEsol} actually solves
\eqref{Eq:GF} for $r=1$ by substitution.
Hence the mean fitness at generation $t$ for $r=1$ is
\begin{equation}
w_t = F_t(e^{s_1},e^{s_2}) = \widetilde G_0(e^{s_1 t})\widetilde G_0(e^{s_2 t}).
\label{Eq:SI}
\end{equation}
One should note that the ansatz \eqref{Eq:part} successfully
gives the exact solution because the homogeneous initial condition 
satisfies \eqref{Eq:part}, but the speed does not depend on the initial 
condition as long as the maximum number of existing mutations at $t=0$ is 
finite.

From Equations \ref{Eq:AI} and \ref{Eq:SI}, we deduce the speed of
adaptation as
\begin{eqnarray}
\label{Eq:vadef}
v_a \equiv v(r=0,N=\infty) = \lim_{t \rightarrow \infty} \frac{\ln G_0(e^{{s_1}t}, e^{{s_2}t})}{t}, \\
v_s \equiv v(r=1,N=\infty) = \lim_{t \rightarrow \infty} \frac{\ln \widetilde G_0(e^{s_1 t})+\ln \widetilde G_0(e^{s_2 t})}{t}, 
\label{Eq:vsdef}
\end{eqnarray}
where subscripts $a$ and $s$ stand for asexuals and (obligate) sexuals,
respectively.
Since the arguments of $G_0$ in \eqref{Eq:vadef} and of $\widetilde G_0$ in 
\eqref{Eq:vsdef} increase exponentially, the speed is fully determined by the 
largest possible fitness effect due to a single mutation event. Thus,
\begin{eqnarray}
v_a &=& \text{Max}\{ n_1 s_1 + n_2 s_2 | g_0(n_1,n_2) \neq 0\},
\label{Eq:speed_inf_a}
\\
v_s &=&   M (s_1 + s_2),
\label{Eq:speed_inf_s}
\end{eqnarray}
where $M$ is the largest possible number of sites mutated at {\it one locus} in
a single mutation event,
\begin{equation}
M = \text{Max}\{ n| \tilde g_0(n) \neq 0\}.
\label{Eq:def_M}
\end{equation}
Since, by definition, $M$ is the maximum of all possible $n_1$ and $n_2$ 
with $g_0(n_1,n_2) \neq 0$, $v_s$ cannot be smaller than
$v_a$.  Thus, sex is at least not detrimental, though it may 
have no effect depending on the form of $g_0$. 
For example, if single mutations occur with probability $U$ and double
mutations involving both loci with probability $U^2$, corresponding to  
\begin{equation}
g_0(0,0) = 1 - U - U^2,\quad g_0(1,0) = g_0(0,1) = \frac{U}{2},\quad g_0(1,1) = U^2,
\label{Eq:mut2}
\end{equation}
then $v_s = v_a =  s_1 + s_2$.
On the other hand, if double mutations are forbidden and 
\begin{equation}
g_0(0,0) = 1 - U,\quad g_0(1,0) = g_0(0,1) = \frac{U}{2}
\label{Eq:App_mut}
\end{equation}
we have $v_s = s_1+s_2> v_a = s_2$ (recall that we assume $s_2 \ge s_1$). 
Hence the effect of sex significantly depends on the
form of $g_0$ in the infinite population limit.
If $s_2 > s_1$ (strict inequality) and if $g_0$ is as in \eqref{Eq:mut},
beneficial mutations occurring at locus 1 do not
contribute to the speed of an infinite asexual population. This can be understood 
in the framework of clonal interference as the `wasting' of weaker beneficial mutations
by the competition with stronger mutations. 

\renewcommand{\theequation}{B\arabic{equation}}
\setcounter{equation}{0}
\section{APPENDIX B: Guess relation in the presence of recombination}
In this APPENDIX, we will show that for evolution on multiplicative, 
non-epistatic fitness landscapes the Guess relation (\eqref{Eq:Guess}) is valid even in the presence of recombination. 

Let $w_i(t)$ be the fitness of the
$i$-th individual at generation $t$, $\bar w(t)$ the mean fitness of the
population, $\bar w(t) = \sum_i w_i(t)/N$, 
and $X_i(t)$ the relative fitness of $i$-th individual, 
$X_i(t) = w_i(t)/\bar w(t)$. We will assume that $X_i(t)$ approaches a well-defined
steady state as $t$ goes to infinity. 
We take each individual to be characterized by a genome 
with $L$ loci, each of which has infinitely many sites. The contribution of a locus
to fitness is denoted by $z_n$ ($n = 1,\ldots,L$) and the fitness of an 
individual with such a genome is $w = \prod_{n=1}^L z_n$. 
In the following,
$z_n$ will be called CF$n$, meaning the Contribution to Fitness of the
$n$th-locus.
If a mutation hits the $n$-th locus, CF$n$ changes from
$z_n$ to $z_n' = z_n v_n$, where $v_n$ is drawn from a 
given probability distribution that may vary from locus to locus but does not 
depend on $z_n$ or the generation.
If $v_n$ is larger (smaller) than 1, the mutation is beneficial (deleterious).
In this APPENDIX, the explicit form of the probability distribution for 
$v_n$ does not need to be specified.

We will use the vector notation $\vec{z} = (z_1,...,z_L)$ for 
fitness vectors with $L$ elements. 
Assume that there are $N$ individuals and the CF$n$ of
individual $i$ is $z_{i,n}$. The corresponding fitness vector 
is denoted by $\vec z_i$. 
At first, we will calculate the expected mean log-fitness at generation
$t+1$ assuming that $X_i(t)$ and $\vec z_i(t)$ are given.

By selection, the probability density that the fitness vector of an offspring
is $\vec z$ is
\begin{equation}
f_s(\vec z) = \sum_{i=1}^N \frac{X_i(t)}{N} \delta(\vec z-\vec z_{i}),
\label{Eq:Guess_sel}
\end{equation}
where $\delta(\vec x)$ is the $L$-dimensional Dirac delta function. 
Let $g(\vec v)$ be the probability density that a
mutation event changes the CF$n$ of an offspring by $v_n$ ($z_n
\rightarrow z_n' = z_n v_n$) for all $n$'s.
Then due to mutation, the expected frequency becomes
\begin{equation}
f_m(\vec z) = \int d\vec v  d\vec z' \delta(\vec z-\vec z'\otimes \vec v)
g(\vec v) f_s(\vec z'),
\label{Eq:Guess_mut}
\end{equation}
where $\vec z'\otimes\vec v$ denotes the vector with elements $z'_n v_n$.

Next we consider recombination.
Let $R(\vec z | \vec z_1,\vec z_2)$ be the probability density that
offspring resulting from the recombination of two parents with
fitness vectors $\vec z_1$ and $\vec z_2$ has fitness $\vec z$.
In general, we can write $R$ in the form
\begin{equation}
R(\vec z | \vec z_1, \vec z_2) = \sum_S p(S) \prod_{n=1}^L
\delta\left (z_n - z_{S(n),n} \right ),
\label{Eq:Guess_R}
\end{equation}
where $S$ runs over all possible outcomes of recombination and $p(S)$ is the
probability of this event. Here $S(n)=1$ (2) if locus $n$ is inherited
from parent 1 (2), hence the total number of 
possible outcomes is $2^L$. Then the final probability density becomes
\begin{equation}
f(\vec z) = \int d\vec z_1 d\vec z_2 R(\vec z | \vec z_1,\vec z_2)
f_m(\vec z_1)f_m(\vec z_2).
\label{Eq:Guess_rec}
\end{equation}

Now we will calculate the expected log-fitness of a randomly chosen individual
at generation $t+1$ for given $w_i(t)$. Since the probability density that a 
randomly chosen individual at generation $t+1$ has fitness vector $\vec z$ is 
$f(\vec z)$ given in \eqref{Eq:Guess_rec} and the corresponding fitness 
is $w = \prod_n z_n$, 
the quantity we want to calculate is
\begin{equation}
I \equiv \int dw \ln w~\text{prob}(w) = \int  d\vec z \ln\left ( \prod_{n=1}^L z_n \right ) f(\vec z),
\label{Eq:I}
\end{equation}
where $\text{prob}(w)$ is the probability density that an individual has fitness $w$ at generation $t+1$ for given $w_i(t)$.
\citeN{G1974A} showed that as long as there is a well-defined steady state the speed $v$ can be calculated as 
\begin{equation}
v = \left \langle \ln \frac{\bar w(t+1)}{\bar w (t)} \right \rangle = \left \langle \ln \frac{{\cal W}_1}{{\cal W}_2} \right \rangle 
= \langle I \rangle - \left \langle \ln {{\cal W}_2} \right \rangle ,
\label{Eq:Guess_v}
\end{equation} 
where ${\cal W}_1$ (${\cal W}_2$) is the fitness of a randomly chosen individual at
generation $t+1$ ($t$) and $\langle \ldots \rangle$ signifies the
average over the steady state distribution.
Loosely speaking, the above relation can be understood as follows:
Since the dynamics is symmetric under permutations of the population index, the steady
state must have this permutation symmetry as well. Hence the expected log-mean fitness 
at steady state should be the same as the expected log-fitness of a 
randomly chosen individual. 
Moreover, at stationarity the speed can be calculated from the difference of log-fitness
between two consecutive generations. Thus \eqref{Eq:Guess_v} follows.

From Equations \ref{Eq:Guess_sel}, \ref{Eq:Guess_mut}, \ref{Eq:Guess_R},
and \ref{Eq:Guess_rec}, we get 
\begin{eqnarray}
I 
&=& \int  d\vec z \ln\left ( \prod_{n=1}^L z_n \right )
\sum_{i,j} \frac{X_i(t) X_j(t)}{N^2} \int d \vec v_i d \vec v_j
R(\vec z | \vec z_i\otimes \vec v_i, \vec z_j\otimes \vec v_j) g(\vec v_i) g(\vec v_j)
\nonumber \\
&=&\sum_{i,j} \frac{X_i(t) X_j(t)}{N^2} \int d \vec v_i d \vec v_j g(\vec v_i) g(\vec v_j)
 \int  d\vec z \ln\left ( \prod_{n=1}^L z_n \right )
R(\vec z | \vec z_i\otimes \vec v_i, \vec z_j\otimes \vec v_j)
\nonumber \\
 &=& \sum_{i,j} \sum_{S} p(S) \frac{X_i(t) X_j(t)}{N^2}
\int d \vec v_i d \vec v_j g(\vec v_i) g(\vec v_j)
\sum_{n=1}^L\left [ \ln (z_{S(n),n}) + \ln (v_{S(n),n}) \right ]
\nonumber \\
&=&\sum_{S} p(S) \sum_{n=1}^L \left (
\sum_{i,j} \frac{w_{i}w_{j}}{N^2}  \ln (z_{S(n),n})
+ \int d \vec v_i d \vec v_j g(\vec v_i) g(\vec v_j)
\ln (v_{S(n),n}) \right ),
\label{Eq:main}
\end{eqnarray}
where $S(n)$ in the subscript of $z$ and $v$ can be either $i$ or $j$. Since
\begin{eqnarray}
\sum_{ij} \frac{X_i(t) X_j(t)}{N^2} \ln (z_{i,n})=
\sum_{ij} \frac{X_i(t) X_j(t)}{N^2} \ln (z_{j,n}) = \sum_i \frac{X_i(t)}{N}
\ln(z_{i,n}),\\
\int d \vec v_i d \vec v_j g(\vec v_i) g(\vec v_j ) \ln v_{i,n}
=\int d \vec v_i d \vec v_j g(\vec v_i) g(\vec v_j ) \ln v_{j,n}
= \int d \vec v g(\vec v) \ln v_{n},
\end{eqnarray}
the summation and integral in the parentheses of the last line of
\eqref{Eq:main}  do not depend on $S$. Due to the normalization $\sum_S p(S) = 1$,
we get
\begin{equation}
I = \sum_i \frac{X_i(t)}{N} \ln (X_i(t) \bar w(t) ) + \langle \ln V \rangle
  = \sum_i \frac{X_i(t)}{N} \ln (X_i(t) ) + \langle \ln V \rangle + \ln \bar w(t),
\end{equation}
where $V$ is the total effect of fitness increase by a single mutation
event, $V = \prod_n v_n$, and $\prod_n z_{i,n} = X_i(t) \bar w(t)$ was used. 
Hence the speed becomes
\begin{equation}
v =\left  \langle I\right \rangle - \left \langle \sum_i \frac{1}{N} \ln w_i\right \rangle
= \left \langle \ln V \right \rangle + \left \langle \frac{1}{N}\sum_i (\chi_i - 1) \ln \chi_i \right \rangle,
\end{equation}
where $\chi_i$ is the relative fitness of the individual $i$ at steady state.
Note that the formula does not depend on the explicit form of the
recombination operator.
If we use \eqref{Eq:mut} for the mutation scheme, $\langle \ln V \rangle = Us$.

The application of the above procedure to the Moran model is
straightforward. 
Hence, the Guess relation is valid for discrete time models regardless
of recombination, once the fitness landscape is multiplicative.
\newpage
\begin{center}
{\huge Supporting Information}
\end{center}
\newpage
\begin{center}
File S1

\verb
FILE_S1.gif
\end{center}
File S1: Animation of a breathing traveling wave in frequency space. $n_1$ and $n_2$ stands for the number of mutations
at locus 1 and locus 2, respectively. Genotypes with frequency larger than $10^{-5}$ are shown.
\end{document}